\documentclass{article}
\usepackage[margin=2.5cm]{geometry}
\usepackage{authblk}
\usepackage{mathtools,amsmath, amssymb, bm, siunitx,dutchcal}
\usepackage[font=small,labelfont=bf]{caption}
\usepackage{subcaption}
\usepackage{enumerate}
\usepackage{algorithm,algpseudocode}
\usepackage{booktabs}
\DeclarePairedDelimiter{\norm}{\lVert}{\rVert_\text{2}}

\usepackage{tikz,pgfplots}

\usepackage{hyperref}
\hypersetup{
	linkcolor  = blue,
	citecolor  = red,
	urlcolor   = blue,
	colorlinks = true,
}

\usepackage[capitalize]{cleveref}
\usepackage[style=authoryear-icomp,uniquelist=false,maxcitenames=1,maxbibnames=99,url=false,hyperref=true,sorting=ynt, isbn=false, eprint=true, doi=true, url=false, backend=biber, giveninits, uniquename=false]{biblatex}
\begin{document}

\title{Efficient computation of states and sensitivities for compound structural optimisation problems using a Linear Dependency Aware Solver (LDAS)}

\author[1]{Stijn Koppen}
\author[1]{Max van der Kolk}
\author[2]{Sanne van den Boom}
\author[1]{Matthijs Langelaar}
\affil[1]{Precision \& Microsystems Engineering, Delft University of Technology, Mekelweg 2, 2628 CD Delft, The Netherlands}
\affil[2]{Structural Dynamics, Netherlands Institute for Applied Scientific Research (TNO), Leeghwaterstraat 44--46, 2628 CA Delft, The Netherlands}
\date{\today}

\twocolumn
\maketitle

\begin{abstract}
    Real-world structural optimisation problems involve multiple loading conditions and design constraints, with responses typically depending on states of discretised governing equations.
Generally, one uses gradient-based nested analysis and design approaches to solve these problems.
Herein, solving both physical and adjoint problems dominates the overall computational effort. Although not commonly detected, real world problems can contain linear dependencies between encountered physical and adjoint loads. Manually keeping track of such dependencies becomes tedious as design problems become increasingly involved.
To detect and exploit such dependencies, this work proposes the use of a Linear Dependency Aware Solver (LDAS), which is able to efficiently detect linear dependencies between all loads to avoid unnecessary solves entirely and automatically. Illustrative examples are provided that demonstrate the need and benefits of using an LDAS, including a run-time experiment.
\end{abstract}

\section{Introduction}
\label{sec:intro}

In structural optimisation, particularly in topology optimisation, the \emph{self-adjoint} compliance minimisation problem is often studied \autocite{Rozvany1989}. As a consequence of self-adjointness, one can obtain design sensitivities for gradient-based optimisation at marginal computational cost. This advantage has likely contributed to the popularity of studying the compliance minimisation problem. However, as \textcite{Rozvany1993} pointed out almost three decades ago: ``Self-adjoint problems, such as design for a single stress, a single compliance or single natural frequency constraint do not represent a real-world situation, because most practical structures are subject to several load conditions and design constraints.'' Almost three decades later, solving large-scale linear problems considering multiple physical loads and a large variety of responses---hereafter denoted by \emph{compound} problems---is becoming increasingly attainable as available computational power increases. However, regardless of available computational power, efficient numerical implementations remain essential.

Typically, finding the state corresponding to a load, \textit{i.e.} the solution to the governing equations, dominates the overall computation time during optimisation. As \textcite{Borrvall2001} report, the computational time of such procedures approaches 97\% for minimum compliance problems considering a single physical load, where computation times increase further when considering compound problems.

Finding a solution to the involved systems of linear equations generally consist of two steps: preprocessing and solving \autocite{Amir2010}. The preprocessing for direct methods requires the (generally expensive) matrix factorization and solving requires finding the exact solution via comparatively inexpensive back-substitutions \autocite{davis2006direct}.  In contrast, iterative methods require the construction of a preconditioner, and they subsequently generate a sequence of approximate solutions until convergence \autocite{Saad2003}.  The relative cost of preconditioner construction  and  the  iterative  solution  process  depends  on  many  factors,  such  as the type of preconditioner and condition number. The preprocessing information can, in many cases, be repeatedly reused, which is particularly advantageous when multiple loads are considered.

Typically, finding the state corresponding to a load, \textit{i.e.} the solution to the governing equations, dominates the overall computation time during optimisation. As \textcite{Borrvall2001} report, the computational time of such procedures approaches 97\% for minimum compliance problems considering a single physical load, where computation times increase further when considering compound problems.

Finding a solution to the involved systems of linear equations generally consist of two steps: preprocessing and solving \autocite{Amir2010}. The preprocessing for direct methods requires the (generally expensive) matrix factorization and solving requires finding the exact solution via comparatively inexpensive back-substitutions \autocite{davis2006direct}.  In contrast, iterative methods require the construction of a preconditioner, and they subsequently generate a sequence of approximate solutions until convergence \autocite{Saad2003}.  The relative cost of preconditioner construction  and  the  iterative  solution  process  depends  on  many  factors,  such  as the type of preconditioner and condition number. The preprocessing information can, in many cases, be repeatedly reused, which is particularly advantageous when multiple loads are considered.

Three strategies can be distinguished to lower the computational effort of solving large-scale linear systems of governing equations in structural optimisation, \textit{i.e.} reduction of 
\begin{enumerate}[\itshape i]
\item the number of design iterations, 
\item the computational effort per solve, and 
\item the number of solves per design iteration. 
\end{enumerate} 
The first technique has shown great potential to reduce computational effort, for instance using advanced sequential approximate optimisation schemes (\textit{e.g.} see \autocite{Bruyneel2002b}). However, these approaches are considered out of scope for this discussion, as they are independent of the presented methodology. 

A common approach to reduce computation \emph{time} per linear solve is to employ parallel computing \autocite{Borrvall2001,Aage2017}, a technique which \emph{distributes} the computational effort.
However, to \emph{reduce} this effort, approximation techniques should be considered, such as approximated reanalysis \autocite{kirsch1991reduced,Amir2015}, iterative solution techniques \autocite{Borrvall2001,Amir2010a,Amir2014}, and approximated model order reduction \autocite{Ma1993, Choi2019}.
Alternatively, static condensation \autocite{Guyan1965, Irons1965} allows for exact model order reduction, decreasing the system dimensionality without loss of information (\textit{e.g.} see \autocite{Yang1996}). 
For a comprehensive review of techniques aiming to decrease the computational effort per solve in context of topology optimisation, the reader is referred to the recent work by \textcite{Mukherjee2021}.

The third category---approaches to reduce the number of solves per design iterations---includes the adjoint sensitivity analysis method itself, for instance when applied to most self-adjoint problems \autocite{Arora1979,Vanderplaats1980,Belegundu1986}. 
For problems considering many physical loads, \textcite{Zhang2020} reduce the number of deterministic loads to a single approximated load using sampling schemes.
Static condensation also allows for considerable reduction of the number of solves per design iteration for problems with multiple partitions, that is the problem involves multiple load cases with different boundary conditions \autocite{koppen2021efficient}.

In this paper, we introduce another method of the third category that reduces the number of solves per design iteration. 
We herein assume linear state-based optimization problems under (quasi-)static loading, which constitutes a large fraction of all problems studied in the topology optimization community \autocite{Bendsoe2004}.
By automatically detecting linear dependencies between physical and adjoint loads, unnecessary solves in compound problems can be avoided entirely. 
For clarity, we distinguish three cases of such linear dependency:
\begin{enumerate}[\itshape i]
\item cases where two physical loads are linearly dependent, from now on referred to as Linearly Dependent Physical-Physical (LDPP) loads,
\item cases where the adjoint load depends linearly on the \emph{corresponding} physical load, as is common in conventional self-adjoint\footnote{It is a common misconception that self-adjoint problems \emph{always} exhibit an LDAP pair, as such problems can (and originally were) often of analytical nature and/or do not require a solve to obtain sensitivities (\textit{e.g.} design for a single natural frequency) \autocite{shield1970optimal,Rozvany1993}. Also, problems that exhibit an LDAP pair are by no definition \textit{per se} self-adjoint (\textit{e.g.} the optimisation for deflection constraints constitutes a \emph{non-self-adjoint problem}, although exhibiting an LDAP pair \autocite{Rozvany1993}).} problems, referred to as Linearly Dependent Adjoint-Physical (LDAP) load pairs, and
\item cases where physical loads or adjoint loads can be written as any linear combination of previously considered physical and/or adjoint loads; that is Mixed Linear Dependency (MLD).
\end{enumerate} 
MLDs also include linear dependencies between adjoint loads as well as between non-corresponding adjoint and physical loads (as well as any linear combination). This is the most general situation, and as such the most difficult to foresee and consider by hand. A Linear Dependency Aware Solver (LDAS) can be employed to automatically detect and exploit these dependencies. 
In this work, we demonstrate the need and benefits of an LDAS in context of gradient-based, structural optimisation for compound problems, and provide one such solver in the form of a simple algorithm to automatically detect and exploit any linear dependence in a (possibly large) set of loads. The focus is on the general case of MLDs. However, due to the generality of the method, it also automatically resolves unnecessary solves in LDPP and LDAP pairs (as well as other combinations). Thus, it is ensured that in each iteration only the minimum number of linear solves is performed. This makes the approach very suitable for implementation in general purpose structural and topology optimization packages.
It should be noted that the presented algorithm does \emph{not} exclude other additional techniques to reduce the computational effort and time, such as parallel computing, approximate modelling, or reduced order techniques, which can be implemented alongside the presented methodology. 

\section{Method}
\label{sec:method}

Consider a general inequality-constrained nonlinear structural optimisation problem
\begin{align}
\label{eq:p}
\begin{aligned}
\underset{\mathbf{s}\in \mathbb{S}^N}{\text{min}} &&& f\left[\mathbf{s}, \mathbf{U}\left[\mathbf{s}\right]\right]\\
\text{s.t.} &&& \mathbf{g}\left[\mathbf{s}, \mathbf{U}\left[\mathbf{s}\right]\right] \leq \mathbf{0}\\
\end{aligned}
\end{align}
with objective $f \in \mathbb{R}$,
$m$ inequality constraints $\mathbf{g} \in \mathbb{R}^{m}$ and $N$ design variables $\mathbf{s} \in \mathbb{S}^{N} \subseteq \mathbb{R}^N$.

\subsection{Response and sensitivity analysis}
The responses (objective and constraint functions) commonly depend on physical states $\mathbf{U} := \left[\mathbf{u}_1 ,\ldots, \mathbf{u}_a\right] \in \mathbb{R}^{n \times a}$, where $n$ is the dimensionality of the discretised governing equations and $a$ the number of states. 
These states implicitly depend on the design variables, \textit{i.e.} $\mathbf{U} = \mathbf{U}\left[\mathbf{s}\right]$. 
We consider a setting in which these physical states are obtained by solving a linear system of discretised governing equations, \textit{i.e.}
\begin{equation}
\label{eq:axb}
\mathbf{K}\left[\mathbf{s}\right] \mathbf{U} = \mathbf{F}\left[\mathbf{s}\right],
\end{equation}
with $\mathbf{F}\left[\mathbf{s}\right]:= \left[\mathbf{f}_1\left[\mathbf{s}\right] ,\ldots, \mathbf{f}_a\left[\mathbf{s}\right]\right] \in \mathbb{R}^{n \times a}$ the physical loads and $\mathbf{K}\left[\mathbf{s}\right] \in \mathbb{R}^{n \times n}$ a design dependent, symmetric, and non-singular system matrix. In the following we assume the system in \cref{eq:axb} constitutes a single partition, thus the physical loads are applied on the system under the same boundary conditions.

In gradient-based optimisation the sensitivities of the responses with respect to the design variables are required to update the design variables.
For structural optimisation problems with a large ratio of number of design variables to number of state-based response functions, commonly the adjoint method is applied to efficiently obtain this sensitivity information \autocite{Arora1979,Vanderplaats1980}.
To this end, consider the augmented response
\begin{equation}
\label{eq:lag}
\mathcal{L}_j\left[\mathbf{s},\mathbf{U}\left[\mathbf{s}\right]\right] = g_j\left[\mathbf{s},\mathbf{U}\left[\mathbf{s}\right]\right] - \bm{\Lambda}_j : \left(\mathbf{K}\left[\mathbf{s}\right]\mathbf{U} - \mathbf{F}\left[\mathbf{s}\right]\right).
\end{equation}
with $\mathbf{\Lambda}_j := \left[\mathbf{\lambda}_{j,1} ,\ldots, \mathbf{\lambda}_{j,a}\right] \in \mathbb{R}^{n \times a}$. Here, a suitable choice of the adjoint states $\mathbf{\Lambda}_j$ can circumvent calculation of the computationally expensive derivative $\frac{\partial \mathbf{U}}{\partial s_k}$ \autocite{Vanderplaats1980}. Doing so, full differentiation of \cref{eq:lag} yields
\begin{equation}
\label{eq:blamb}
\frac{\text{d} \mathcal{L}_j}{\text{d} s_k} = \frac{\partial g_j}{\partial s_k} - \bm{\Lambda}_j : \left( \frac{\partial \mathbf{K}}{\partial s_k} \mathbf{U}\right),
\end{equation}
with
\begin{equation}
\label{eq:alamb}
\mathbf{K}\left[\mathbf{s}\right] \bm{\Lambda}_j = \frac{\partial g_j}{\partial \mathbf{U}},
\end{equation} where $\frac{\partial g_j}{\partial \mathbf{U}}$ is referred to as the adjoint loads of response $g_j$.

Each of the physical and adjoint loads can be linearly dependent with respect to any combination of previously considered loads, and thus can be reconstructed as their linear combination. Exploiting possible linear dependence can significantly reduce the overall cost required to find all states. Consider a set of $a$ loads, of which $b$ are linearly independent, then the computational effort scales roughly with $\frac{b}{a}$, as only $b$ solves are required to reconstruct all states. To avoid unnecessarily solving \cref{eq:axb,eq:alamb} for linear dependent loads we propose
\begin{enumerate}[\itshape i]
\item to compute each load's dependency on previous loads, and
\item to keep track of the states corresponding to linearly independent loads.
\end{enumerate}
A variety of possible methods exist to check for linear dependency and necessary bookkeeping.
In what follows, we consider one such algorithm that detects linear dependencies and builds orthogonal bases of linear independent loads and their corresponding states.

\subsection{Orthogonalisation and reconstruction}\label{sec:ortho}
Consider the non-empty orthogonal bases of loads $\mathcal{F}$ and states $\mathcal{U}$ of length $c$.
One can investigate the linear dependency of a load $\mathbf{f}$ (\textit{e.g.} a physical load $\mathbf{f}$ or adjoint load $\frac{\partial g}{\partial \mathbf{u}}$) with respect to $\mathcal{F}$ by applying the last step of the well known Gram-Schmidt orthogonalisation procedure\footnote{Although the method is named after J{\o}rgen Pedersen Gram and Erhard Schmidt,  Pierre-Simon Laplace had been familiar with it before, see \autocite{leon2013gram}.}  \autocite{laplace1820theorie,gram1883ueber,schmidt1907theorie}.
The residual $\mathbf{r}$ is obtained via
\begin{equation}
\label{eq:gs}
\mathbf{r} := \mathbf{f} - \sum_{i=1}^{c} \alpha_i \mathcal{F}_i, \quad \text{with} \quad \alpha_i = \frac{\mathcal{F}_i\cdot \mathbf{f}}{\mathcal{F}_i \cdot \mathcal{F}_i},
\end{equation}
with $\mathcal{F}_i$ the \emph{i}th load in $\mathcal{F}$. A possible implementation is given by the pseudo-code \cref{alg:gs}.

\begin{algorithm}
\caption{Gram-Schmidt orthogonalisation}\label{alg:gs}
\begin{algorithmic}[1]
\Function{\textbf{GSO}}{$\mathbf{f},\mathcal{F}$}
\State $\bm{\alpha}$ = []
\State $\mathbf{r}$ = \textbf{copy}$(\mathbf{f})$
\For{$\mathcal{f}$ in $\mathcal{F}$}
\State $\alpha = (\mathbf{r} \cdot \mathcal{f})/(\mathcal{f} \cdot \mathcal{f})$
\State $\mathbf{r} \mathrel{-}= \alpha \mathcal{f}$
\State $\bm{\alpha}$.append$(\alpha)$
\EndFor
\State \Return $(\bm{\alpha}, \mathbf{r})$
\EndFunction
\end{algorithmic}
\end{algorithm}

If the norm of the residual $\mathbf{r}$ is zero, then $\mathbf{f}$ is linearly dependent with respect to basis $\mathcal{F}$.
As a result, the corresponding state $\mathbf{u}$ (or adjoint state $\bm{\lambda}$) is linearly dependent on basis $\mathcal{U}$. Thus, the state $\mathbf{u}$ may be reconstructed via
\begin{equation}
\label{eq:recon}
\mathbf{u} = \sum_{i=1}^{c} \alpha_i \mathcal{U}_i.
\end{equation}
As such, solving the governing equations for $\mathbf{f}$ can be omitted.
However, if the norm of the residual vector $\mathbf{r}$ is nonzero, $\mathbf{f}$ is linearly \textit{independent} with respect to basis $\mathcal{F}$ and the expensive solve cannot be avoided.

We solve for the state $\mathbf{v}$ corresponding to residual load $\mathbf{r}$ defined by
\begin{equation}\label{eq:solve}
\mathbf{K}\left[\mathbf{s}\right]\mathbf{v} = \mathbf{r}.
\end{equation}
Subsequently load $\mathbf{r}$ and state $\mathbf{v}$ are added to bases $\mathcal{F}$ and $\mathcal{U}$, respectively.
Since $\mathbf{r}$ is orthogonal with respect to basis $\mathcal{F}$, so is $\mathbf{v}$ to $\mathcal{U}$.
As a result, both enriched bases $\mathcal{F}$ and $\mathcal{U}$ remain orthogonal.
The state $\mathbf{u}$ is then reconstructed from \cref{eq:gs} and \cref{eq:recon}. The above procedure can be repeated using the enriched bases, as defined in \cref{alg:axb}.

\begin{algorithm}
\caption{Linear Dependency Aware Solver}
\label{alg:axb}
\begin{algorithmic}[1]
\Function{\textbf{LDAS}}{$\mathbf{K},\mathbf{F},\mathbf{U},\mathcal{F},\mathcal{U},\varepsilon=10^{-6}$}
\For{$(i, \mathbf{f})$ in enumerate$(\mathbf{F})$}
\State $(\bm{\alpha}, \mathbf{r})$ = \textbf{GSO}$(\mathbf{f}, \mathcal{F})$
\If{$\norm{\mathbf{r}} > \varepsilon $}
\State $\mathcal{F}$.append$(\mathbf{r})$
\State $\mathcal{U}$.append$(\textbf{solve}(\mathbf{K}, \mathbf{r}))$
\State $\bm{\alpha}$.append$(1)$
\EndIf
\State $\mathbf{U}[i] \mathrel{+}= \bm{\alpha} \cdot \mathcal{U}$
\EndFor
\State \Return $\mathbf{U}$
\EndFunction
\end{algorithmic}
\end{algorithm}

Although \cref{alg:axb} introduces additional computational operations, \textit{i.e.} computing vector norms and orthogonality coefficients, their computational cost is typically negligible compared to the costs of solving a system of equations, as illustrated in \cref{sec:numerical}. Furthermore, these operations do not change when considering distributed-memory parallelism. Alternatively, for loads that do not depend on the states, it is possible to rearrange \cref{alg:axb} to determine all the independent loads first and evaluate their solutions in parallel afterwards.

\section{Analytical example}
\label{sec:example}

Compound problems may appear in any real-world problem, modelled by (a sequence of) linear governing equations. 
Typical examples of compound problems are formulations with multiple loading conditions and multiple response functions in which the degrees of freedom of (some of) the loads coincide with (some of) the degrees of freedom that define the response functions.
One may think for example of the design of a structure with multiple critical loading conditions, where the displacements of a loading condition are measured at the same degrees of freedom where the loads are applied at another loading condition.
A direct example of this are multi-input-multi-output compliant mechanisms, see \textit{e.g.} \autocite{Frecker1999} or \autocite{Liu2009}. 
The problem formulation of such mechanisms includes multiple physical loads \emph{and} responses, all applied to, or dependent on, the input and output degrees of freedom of the mechanism. 
As a result, MLD is commonly present, however, generally remains unnoticed.
To clarify the cases in which one might encounter linear dependency, we here exemplify the three different types of unnecessary solves, as introduced in \cref{sec:intro}. 

\begin{figure}
\centering
\def\svgwidth{0.45\textwidth}
\begingroup%
  \makeatletter%
  \providecommand\color[2][]{%
    \errmessage{(Inkscape) Color is used for the text in Inkscape, but the package 'color.sty' is not loaded}%
    \renewcommand\color[2][]{}%
  }%
  \providecommand\transparent[1]{%
    \errmessage{(Inkscape) Transparency is used (non-zero) for the text in Inkscape, but the package 'transparent.sty' is not loaded}%
    \renewcommand\transparent[1]{}%
  }%
  \providecommand\rotatebox[2]{#2}%
  \newcommand*\fsize{\dimexpr\f@size pt\relax}%
  \newcommand*\lineheight[1]{\fontsize{\fsize}{#1\fsize}\selectfont}%
  \ifx\svgwidth\undefined%
    \setlength{\unitlength}{272.7999932bp}%
    \ifx\svgscale\undefined%
      \relax%
    \else%
      \setlength{\unitlength}{\unitlength * \real{\svgscale}}%
    \fi%
  \else%
    \setlength{\unitlength}{\svgwidth}%
  \fi%
  \global\let\svgwidth\undefined%
  \global\let\svgscale\undefined%
  \makeatother%
  \begin{picture}(1,0.23753666)%
    \lineheight{1}%
    \setlength\tabcolsep{0pt}%
    \put(0,0){\includegraphics[width=\unitlength,page=1]{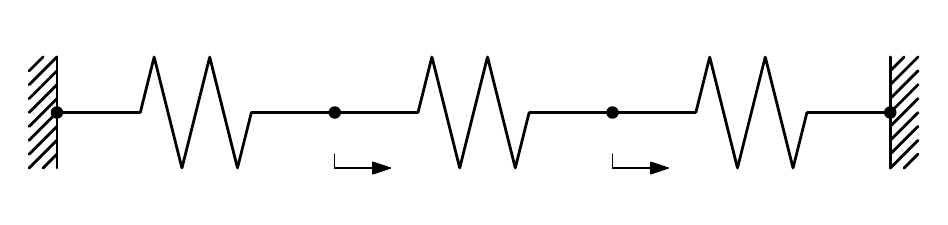}}%
    \put(0.64662757,0.15175953){\color[rgb]{0,0,0}\makebox(0,0)[lt]{\lineheight{1.25}\smash{\begin{tabular}[t]{l}$u_2$\end{tabular}}}}%
    \put(0.35337243,0.15175953){\color[rgb]{0,0,0}\makebox(0,0)[lt]{\lineheight{1.25}\smash{\begin{tabular}[t]{l}$u_1$\end{tabular}}}}%
  \end{picture}%
\endgroup%

\caption{One-dimensional two degrees of freedom compliant mechanism model.}
\label{fig:spring}
\end{figure}

\subsection{Problem formulation}
Consider the two degrees of freedom spring model as depicted in \cref{fig:spring}.
Note that this example---after applying static condensation---can \textit{exactly} represent any single-input-single-output compliant mechanism, see \textit{e.g.} \autocite{Wang2009,Hasse2017}. 
This two degrees of freedom example is therefore fully representative of large-scale linear problems considering multiple physical loads and responses, while better suited to illustrate the proposed method.

\subsection{Forward analysis}
The physical and adjoint states can be obtained by solving the design-dependent discretised governing equations following \cref{eq:axb,eq:alamb}.
A set of the following three physical loads is considered:
\begin{equation}
\label{eq:B}
\mathbf{F} = \begin{bmatrix} \begin{bmatrix} 1 \\ 0\end{bmatrix} \begin{bmatrix} 1 \\ 2\end{bmatrix} \begin{bmatrix} 4 \\ 4\end{bmatrix}\end{bmatrix}.
\end{equation}
The first residual \textit{by definition} equals the first load, that is $\mathbf{r}_1 = \mathbf{f}_1$. As a result, the state $\mathbf{v}_1 = \mathbf{u}_1$.
Since the basis is initially empty when this load is considered, the resulting load and state are directly added to corresponding bases.
The second residual is calculated via \cref{eq:gs}, that is
\begin{equation}
\mathbf{r}_2 = \mathbf{f}_2 - \alpha_1 \mathcal{F}_1 = \begin{bmatrix} 0 \\ 2 \end{bmatrix}.
\end{equation}
Since $\mathbf{r}_2$ is non-zero, the first and second physical loads are linearly independent.
The corresponding physical state $\mathbf{v}_2$ is obtained by solving for the non-zero load $\mathbf{r}_2$ via \cref{eq:solve}.
As a result the following bases, consisting of orthogonal vectors, are obtained after solving for the first two loads:
\begin{equation}
\mathcal{F} = \left[\mathbf{f}_1, \mathbf{r}_2\right] \quad \text{and} \quad \mathcal{U} = \left[\mathbf{u}_1, \mathbf{v}_2\right].
\end{equation}
The second physical state is now reconstructed following \cref{eq:recon} and reads
\begin{equation}
\mathbf{u}_2 = \alpha_1 \mathcal{U}_1 + \mathbf{v}_2= \mathbf{u}_1 + \mathbf{v}_2.
\end{equation}

The third physical load can be written as a linear combination of the current orthogonal basis $\mathcal{F}$, resulting in a zero residual load $\mathbf{r}_3 = \mathbf{0}$, this is thus an LDPP case.
Thus the basis $\mathcal{U}$ can be used to reconstruct the third physical state without an additional solve as in \cref{eq:recon}, \textit{i.e.}
\begin{equation}
\mathbf{u}_3 = \alpha_1\mathcal{U}_1 + \alpha_2 \mathcal{U}_2 = 4\mathbf{u}_1 + 2\mathbf{v}_2.
\end{equation}

\begin{table*}[]
\centering
\begin{tabular}{p{1.5cm}p{1.5cm}|p{1.5cm}p{1.5cm}p{1.5cm}p{1.5cm}p{1.5cm}p{1.5cm}}
\toprule
\multicolumn{2}{c|}{$\mathcal{F}$}&
\multicolumn{6}{c}{Loads}\\
\midrule
$\mathbf{r}_1 = \mathbf{f}_1$ & $\mathbf{r}_2$ & $\mathbf{f}_1$ & $\mathbf{f}_2$ & $\mathbf{f}_3$ & $\frac{\partial g_1}{\partial \mathbf{u}_2}$ & $\frac{\partial g_2}{\partial \mathbf{u}_1}$ &$ \frac{\partial g_2}{\partial \mathbf{u}_3}$\\
\midrule
$\begin{bmatrix} 1 \\ 0\end{bmatrix}$ & $\begin{bmatrix} 0 \\ 2\end{bmatrix} $&
$\begin{bmatrix} 1 \\ 0\end{bmatrix}$ & $\begin{bmatrix} 1 \\ 2\end{bmatrix} $&$\begin{bmatrix} 4 \\ 4\end{bmatrix}$ &$ \begin{bmatrix} \frac{1}{2}\\ 1 \end{bmatrix} $& $\begin{bmatrix} 2 \\ 1 \end{bmatrix} $&$ \begin{bmatrix} 1 \\ 3 \end{bmatrix}$\\
\midrule
& &  $\mathbf{r}_1$ & $\mathbf{r}_1 + \mathbf{r}_2$ & $4\mathbf{r}_1 + 2\mathbf{r}_2$ & $\frac{1}{2}\mathbf{r}_1 + \frac{1}{2}\mathbf{r}_2$&$2\mathbf{r}_1 + \frac{1}{2}\mathbf{r}_2$ & $\mathbf{r}_1 + \frac{3}{2}\mathbf{r}_2$\\
\rule{0pt}{3ex}\\
\multicolumn{2}{c|}{$\mathcal{U}$}&
\multicolumn{6}{c}{States}\\
\midrule
$\mathbf{v}_1 = \mathbf{u}_1$ & $\mathbf{v}_2$ &
$\mathbf{u}_1$ & $\mathbf{u}_2$ & $\mathbf{u}_3$ & $\bm{\lambda}_{1,2}$ & $\bm{\lambda}_{2,1}$ & $\bm{\lambda}_{2,3}$\\
\midrule
&&$\mathbf{v}_1$ & $\mathbf{v}_1 + \mathbf{v}_2$ & $4\mathbf{v}_1 + 2\mathbf{v}_2$ & $\frac{1}{2}\mathbf{v}_1 + \frac{1}{2}\mathbf{v}_2$&$2\mathbf{v}_1 + \frac{1}{2}\mathbf{v}_2$ & $\mathbf{v}_1 + \frac{3}{2}\mathbf{v}_2$\\
\bottomrule
\end{tabular}
\caption{Overview of both physical and adjoint loads and states, as well as the orthogonal bases encountered in the illustrative example presented in \cref{fig:spring}. The right-hand side displays the load and states vectors expressed as linear combinations of the corresponding bases given on the left-hand side.}
\label{tab:spring}
\end{table*}

\subsection{Sensitivity analysis}
Now consider a response function $g_1\left[\mathbf{u}_2\right]$ that is a measure for the strain energy due to load $\mathbf{f}_2$, \textit{i.e.}
\begin{equation}
g_1\left[\mathbf{u}_2\right] = \frac{1}{2}\mathbf{f}_2 \cdot \mathbf{u}_2.
\end{equation}
The second adjoint load for this response is linearly dependent on the corresponding physical load $\mathbf{f}_2$ as
\begin{equation}
\frac{\partial g_1}{\partial \mathbf{u}_2} = \frac{1}{2} \mathbf{f}_2,
\end{equation}
thus this is an LDAP pair, and consequently $\mathbf{r}_4 = \mathbf{0}$.
As a result, one can use the basis $\mathcal{U}$ to reconstruct the second adjoint state, which yields
\begin{equation}
\bm{\lambda}_{1,2} = \frac{1}{2}\mathbf{u}_2 = \alpha_1\mathcal{U}_1 + \alpha_2 \mathcal{U}_2 = \frac{1}{2}\mathbf{u}_1 + \frac{1}{2}\mathbf{v}_2,
\end{equation}
with $\bm{\lambda}_{j,i}$ the adjoint state of response $j$ with respect to state $i$.
Note that both the first and third adjoint loads of this response, that is $\frac{\partial g_1}{\partial \mathbf{u}_1}$ and $\frac{\partial g_1}{\partial \mathbf{u}_3}$ are zero, and thus so are $\bm{\lambda}_{1,1}$ and $\bm{\lambda}_{1,3}$.

Finally consider a (fictitious) response function $g_2\left[\mathbf{u}_1, \mathbf{u}_3\right]$ that depends on both degrees of freedom of the first state \textit{and} third state via
\begin{equation}
g_2\left[\mathbf{u}_1,\mathbf{u}_3\right] = \begin{bmatrix} 2 \\ 1 \end{bmatrix} \cdot \mathbf{u}_1 + \begin{bmatrix} 1 \\ 3 \end{bmatrix} \cdot \mathbf{u}_3.
\end{equation}
The adjoint loads for this response function can be written as
\begin{equation}
\begin{aligned}
\frac{\partial g_2}{\partial \mathbf{u}_1} = \begin{bmatrix} 2 \\ 1 \end{bmatrix} &=  2 \mathbf{f}_1 + \frac{1}{2}\mathbf{r}_2  \quad \text{and} \\
\frac{\partial g_2}{\partial \mathbf{u}_3} = \begin{bmatrix} 1 \\ 3 \end{bmatrix} &= \mathbf{f}_1 + \frac{3}{2}\mathbf{r}_2.
\end{aligned}
\end{equation}
Note that both adjoint loads are linearly dependent on \textit{a combination of} previously considered loads, \textit{i.e.} an MLD.
In this case, the adjoint loads are both linearly dependent on both loads in basis $\mathcal{F}$.
As a result, one may again use the states in $\mathcal{U}$ to reconstruct the adjoint states via
\begin{equation}
\bm{\lambda}_{2,1} = 2 \mathbf{u}_1 + \frac{1}{2}\mathbf{v}_2 \quad \text{and} \quad \bm{\lambda}_{2,3} = \mathbf{u}_1 + \frac{3}{2}\mathbf{v}_2.
\end{equation}

The loads, states, and bases of this example are summarized in Table~\ref{tab:spring}. When all loads (physical and adjoint) are considered, a total of 6 solves are required. If both LDPPs and LDAP pairs are taken into account, only 3 solves are needed. Finally, considering MLDs (and thus also LDPPs and LDAP pairs), only 2 solves are required. 
Although the presented example is simplified, more complex MLDs do appear in large-scale compound problems, as will be demonstrated in \cref{sec:numerical}.

\section{Numerical example}
\label{sec:numerical}

To show the benefits of the proposed method, we consider as illustrative numerical example the topology optimization of a planar, multiple degree-of-freedom micro-mechanism for use, for example, as analog gate in a mechanical computer \autocite{Larsen1997}. Note that the focus here is not on the optimization (problem formulation) of the micro-mechanism, but on demonstrating the numerical benefits of an LDAS.

\subsection{Problem formulation}
\label{sec:probform}
\begin{figure*}
\centering
\begin{subfigure}[t]{0.3\textwidth}
\centering
\def\svgwidth{1\textwidth}
\begingroup%
  \makeatletter%
  \providecommand\color[2][]{%
    \errmessage{(Inkscape) Color is used for the text in Inkscape, but the package 'color.sty' is not loaded}%
    \renewcommand\color[2][]{}%
  }%
  \providecommand\transparent[1]{%
    \errmessage{(Inkscape) Transparency is used (non-zero) for the text in Inkscape, but the package 'transparent.sty' is not loaded}%
    \renewcommand\transparent[1]{}%
  }%
  \providecommand\rotatebox[2]{#2}%
  \newcommand*\fsize{\dimexpr\f@size pt\relax}%
  \newcommand*\lineheight[1]{\fontsize{\fsize}{#1\fsize}\selectfont}%
  \ifx\svgwidth\undefined%
    \setlength{\unitlength}{176.3999956bp}%
    \ifx\svgscale\undefined%
      \relax%
    \else%
      \setlength{\unitlength}{\unitlength * \real{\svgscale}}%
    \fi%
  \else%
    \setlength{\unitlength}{\svgwidth}%
  \fi%
  \global\let\svgwidth\undefined%
  \global\let\svgscale\undefined%
  \makeatother%
  \begin{picture}(1,1)%
    \lineheight{1}%
    \setlength\tabcolsep{0pt}%
    \put(0,0){\includegraphics[width=\unitlength,page=1]{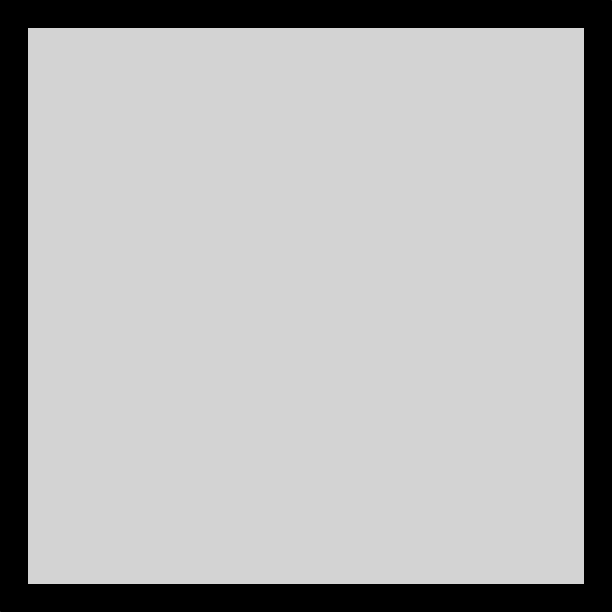}}%
    \put(0.37979136,0.24891991){\makebox(0,0)[lt]{\lineheight{1.25}\smash{\begin{tabular}[t]{l}$5$\end{tabular}}}}%
    \put(0.83279782,0.25030358){\makebox(0,0)[lt]{\lineheight{1.25}\smash{\begin{tabular}[t]{l}$7$\end{tabular}}}}%
    \put(0.74721193,0.33558464){\makebox(0,0)[lt]{\lineheight{1.25}\smash{\begin{tabular}[t]{l}$8$\end{tabular}}}}%
    \put(0,0){\includegraphics[width=\unitlength,page=2]{designprobb.pdf}}%
    \put(0.38053423,0.79346527){\makebox(0,0)[lt]{\lineheight{1.25}\smash{\begin{tabular}[t]{l}$1$\end{tabular}}}}%
    \put(0.29350438,0.87971375){\makebox(0,0)[lt]{\lineheight{1.25}\smash{\begin{tabular}[t]{l}$2$\end{tabular}}}}%
    \put(0.83318033,0.79570149){\makebox(0,0)[lt]{\lineheight{1.25}\smash{\begin{tabular}[t]{l}$3$\end{tabular}}}}%
    \put(0.74743078,0.87960546){\makebox(0,0)[lt]{\lineheight{1.25}\smash{\begin{tabular}[t]{l}$4$\end{tabular}}}}%
    \put(0.29370005,0.33595072){\makebox(0,0)[lt]{\lineheight{1.25}\smash{\begin{tabular}[t]{l}$6$\end{tabular}}}}%
  \end{picture}%
\endgroup%

\subcaption{Initial design and degree of freedom numbering. The intended kinematic degrees of freedom are highlighted using colours; (i) motion from input DOF 6 to output DOF 4, and (ii) motion from input DOF 8 to output DOF 2.}
\label{fig:designprob}
\end{subfigure}
~ 
\begin{subfigure}[t]{0.3\textwidth}
\centering
\includegraphics[width=\textwidth]{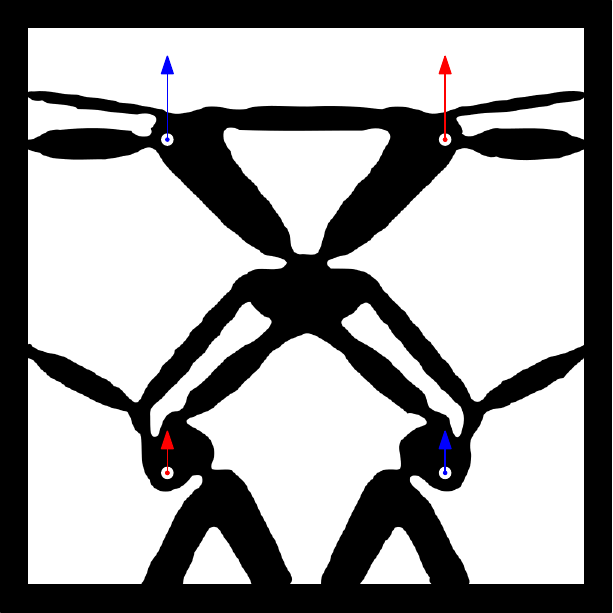}
\subcaption{Final (post-processed) material distribution as obtained from the optimization. Arrow lengths indicate the transmissions between input and output ($J_1 = J_2 = 2$). All other DOFs of interest satisfy the maximum 0.1\% parasitic motion. }
\label{fig:result}
\end{subfigure}
~
\begin{subfigure}[t]{0.3\textwidth}
\centering
\includegraphics[width=\textwidth]{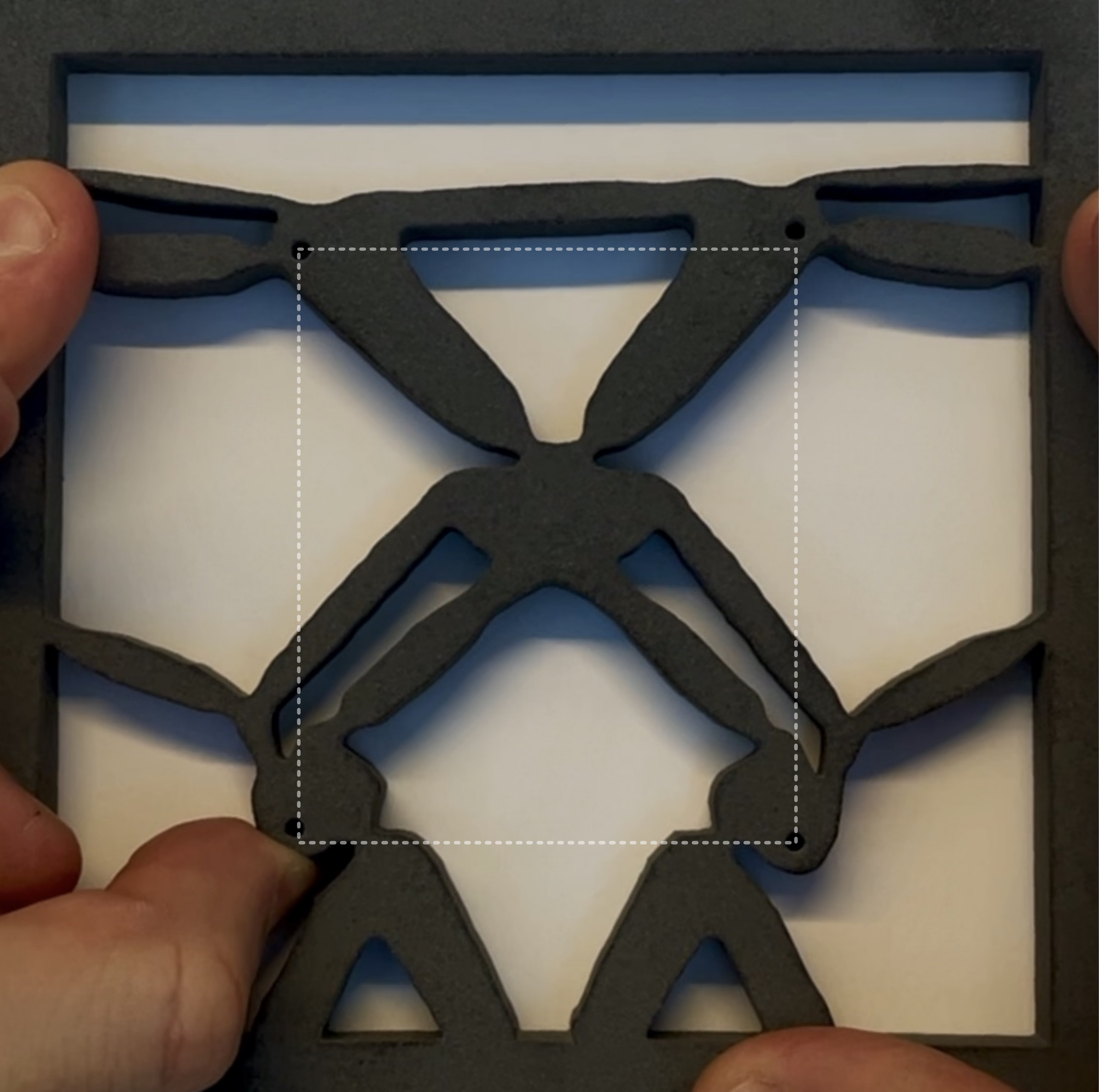}
\subcaption{Prototype design in deformed configuration. The corners of the dashed line indicate the position of the DOFs of interest in undeformed configuration. A movie of the prototype, available as supplementary material (\cref{sec:ror}), demonstrates its functionality.}
\label{fig:proto}
\end{subfigure}
\label{fig:mimo}
\caption{Design of a planar, decoupled multiple degrees of freedom compliant mechanism as described in \cref{sec:probform}. From left to right: (a) the initial design with the four points of interest each with two degrees of freedom ($u_x$, $u_y$), (b) the topology as obtained from the optimization, and (c) a prototype model in deformed configuration.}
\end{figure*}

Consider the design problem depicted in \cref{fig:designprob}. The domain consist of four points of interest, each consisting of two Degrees Of Freedom (DOFs), $u_x$ and $u_y$, respectively. 
The target is to design a monolithic compliant mechanism that \emph{doubles} a unit input motion at DOF 6 to the output motion at DOF 4 \emph{and} a unit input motion at DOF 8 to an equivalent magnified output motion at DOF 2. Thus we consider two independent \emph{kinematic} DOFs.
Furthermore, we also consider parasitic motion, input coupling and output coupling: all remaining DOFs---apart from the intended input and output---are restricted to displace a maximum of 0.1\% of the input motion. 

The force-paths have to cross, making this a challenging problem that is not necessarily intuitive to engineers. Therefore we solve this problem using topology optimization \autocite{Bendsoe2004}. 
We consider the following compound topology optimisation problem formulation\footnote{We do not claim this formulation is (best) suited for the considered problem, we merely employ this formulation for demonstration of the proposed method.}:
\begin{align}
\begin{aligned}
\underset{\mathbf{s}\in \mathbb{S}^N}{\text{min}}\\
f\left[\mathbf{s}\right]:& \quad \sum_j \mathcal{E}_j\left[\mathbf{u}_j\left[\mathbf{s}\right]\right] \quad \forall ~ j \in  \left\{1, 3, 5, 7\right\}\\
\text{s.t.}\\
g^\text{v}\left[\mathbf{s}\right]:&\quad  \sum_k^N s_k \leq N \overline{v}\\
g_{j,j}^\text{in}\left[\mathbf{s}\right] :&\quad  u_{j,j}\left[\mathbf{u}_j\left[\mathbf{s}\right]\right] \geq u_\text{in} \quad \forall ~ j \in \left\{6,8\right\}\\
g_{i,j}^\text{ct}\left[\mathbf{s}\right]:&\quad  u_{i,j}\left[\mathbf{u}_j\left[\mathbf{s}\right]\right] \leq  u_\text{ct} \\ 
&\quad -u_{i,j}\left[\mathbf{u}_j\left[\mathbf{s}\right]\right] \leq  u_\text{ct} \\ 
& \quad \quad \forall~ i,j \in \begin{cases}
\left\{1,2,3,5,7,8\right\},\left\{6\right\}\\
\left\{1,3,4,5,6,7\right\},\left\{8\right\}
\end{cases}\\
g_{i,j}^\text{t}\left[\mathbf{s}\right]:&\quad  J_k u_{i,j}\left[\mathbf{u}_j\left[\mathbf{s}\right]\right] - u_{j,j}\left[\mathbf{u}_j\left[\mathbf{s}\right]\right] \leq u_\text{t} \\
&\quad u_{j,j}\left[\mathbf{u}_j\left[\mathbf{s}\right]\right]-J_k  u_{i,j}\left[\mathbf{u}_j\left[\mathbf{s}\right]\right] \leq u_\text{t}\\
&\quad \quad \forall~ i,j \in \begin{cases}
\left\{4\right\},\left\{6\right\}\\
\left\{2\right\},\left\{8\right\}
\end{cases}
\end{aligned}
\label{eq:problem}
\end{align}
The objective is to minimize the strain energy $\mathcal{E}_j$ by finding design variables $s_k$ that are bounded by $\mathbb{S} = \left\{s \in \mathbb{R} ~ \vert ~ 0 \leq s \leq 1\right\}$.
Constraint $g^\text{v}\left[\mathbf{s}\right]$ limits the maximum material usage by fraction $\overline{v}$. The other constraints enforce a minimum displacement at the input DOFs ($g_{j,j}^\text{in}$), limit cross talk ($g_{i,j}^\text{ct}$), and enforce the transmission between input and output displacements ($g_{i,j}^\text{t}$). In the next subsection these constraints will be further explained.

This problem formulation consists of common, well-documented response functions; an extensive description thereof, as well as corresponding sensitivity analysis, is therefore omitted.
For in-depth discussion on the design of compliant mechanisms using topology optimisation, the reader is referred to earlier works, such as 
\autocite{Ananthasuresh1994,Frecker1997,Sigmund1997b,Sigmund2001} and the review of \textcite{Cao2013} and references therein. For works regarding multiple degrees of freedom systems, the works by \autocite{Frecker1999,Zhan2010,Alonso2014,Zhu2018b,koppen2021simple} may be consulted.

The proposed compound topology optimization problem \cref{eq:problem} was discretized using 200 by 200 finite elements (and design variables) and converged in 58 design iterations using the method of moving asymptotes \autocite{Svanberg1987}. To eliminate modeling artifacts, the design variable field is blurred using a linear convolution operator with filter radius of two elements \autocite{Bruns2001}. A solution, post-processed using bitmap and simplification, is shown in \cref{fig:result}. Note the presence of rigid bodies and hinges, and their location and connections. The resulting deformation and displacements of the DOFs of interest for one of the use-cases is displayed by the prototype in \cref{fig:proto}. A movie of the prototype---available as supplementary material and provided on Github (\cref{sec:ror})---demonstrates that the intended functionality has been achieved.

\subsection{Problem analysis}
\label{sec:probanalysis}
Let us analyse the properties of this optimization problem in light of the proposed method, with a specific emphasis on the required number of systems of equations that are to be solved.

\subsubsection*{Forward analysis}
The objective function $f\left[\mathbf{s}\right]$ is a summation of strain energies, obtained by analysing the deformed structure under a unit load at DOFs $\{1, 3, 5, 7\}$. The internal strain energy corresponding to each displacement field $\mathbf{u}_j$ reads as
\begin{equation}
\mathcal{E}_j = \frac{1}{2} \mathbf{u}_j \cdot \mathbf{K}\left[\mathbf{s}\right] \mathbf{u}_j,
\end{equation}
where $\mathbf{u}_j$ is found by solving the system of equations
\begin{equation}
\label{eq:kuf}
\mathbf{K}\left[\mathbf{s}\right] \mathbf{u}_j = \mathbf{f}_j,
\end{equation}
with $\mathbf{f}_j$ the unit load vector that contains zeros at all entries except at DOF $j$ of interest. To evaluate the objective function, the system of equations (\cref{eq:kuf}) needs to be solved repeatedly, since the four physical loads are linearly independent. 
By minimising these strain energy terms the motion corresponding to these DOFs is restricted in the resulting structure. That is, none of the points of interest can move in $x$-direction. 

Constraints $g_{j,j}^\text{in}\left[\mathbf{s}\right]$ are required to enforce a minimum displacement $u_\text{in}$ at $u_{j,j}$ with $j$ the DOFs of interest 6 and 8, requiring two additional solves. Note, $u_{i,j}$
denotes the displacement at DOF $i$ due to a unit load at DOF $j$. One may observe that the remaining displacement-based constraints are only dependent on $\mathbf{u}_6$ and $\mathbf{u}_8$.
Since these were previously evaluated to determine $g_{j,j}^\text{in}\left[\mathbf{s}\right]$, inspection shows that \emph{no} additional solves are required for the forward analysis.

Constraints $g_{i,j}^\text{ct}\left[\mathbf{s}\right]$ are imposed to limit the crosstalk (parasitic motion) $u_{i,j}$ of DOFs $\{1, 2, 3, 5, 7, 8\}$ due to a unit load at DOF 6 \emph{and} the motion of DOFs $\{1, 3, 4, 5, 6, 7\}$ due to a unit load at DOF 8 from below by $-u_\text{ct}$ and from above by $u_\text{ct}$. 
The number of crosstalk constraints is found by multiplying two kinematic DOF, six constraints per kinematic DOF, and two bounds per constraint, resulting in 24 constraint functions.

Constraints $g_{i,j}^\text{t}\left[\mathbf{s}\right]$ enforce a desired input-output transmission $J_k$ for kinematic DOF $k$ with a maximum transmission deviation of $u_\text{t}$. This introduces four constraints, as each constraint is bound from below \emph{and} above.

All response functions combined require 32 response functions to be evaluated for this optimisation problem, which are fully resolved by performing a total of \emph{six} solves (four for the objective and two for $g_{j,j}^\text{in}\left[\mathbf{s}\right]$). 

\subsubsection*{Sensitivity analysis}
To obtain the sensitivities of the responses with respect to the design variables one generally loops over the responses and \emph{consecutively} calculates the corresponding sensitivities. 

For the considered problem, the adjoint loads of the objective are linearly dependent on \emph{corresponding} physical loads, i.e. they form four LDAP pairs. In this case $\frac{\partial \mathcal{E}_j}{\partial \mathbf{u}_j} = \mathbf{f}_j$, and thus $\bm{\lambda}_{j,j} = \mathbf{u}_j$. Thus, to obtain the sensitivities of the objective no additional solves are required.

The adjoint loads corresponding to $g_{j,j}^\text{in}\left[\mathbf{s}\right]$ read
\begin{equation}
\label{eq:dbc}
\frac{\partial g_{j,j}^\text{in}\left[\mathbf{s}\right]}{\partial \mathbf{u}_j} = \frac{1}{u_\text{in}} \mathbf{l}_j,
\end{equation}
which can be written as a linear combination of the physical loads $\mathbf{f}_6$ and $\mathbf{f}_8$ previously considered to evaluate $g_{j,j}^\text{in}\left[\mathbf{s}\right]$.

The sensitivities of the crosstalk constraints $g_{i,j}^\text{ct}\left[\mathbf{s}\right]$ exhibit MLDs. 
Furthermore, for $i = \{1,3,5,7\}$ and $j = \{6,8\}$ the following holds
\begin{equation}
\label{eq:dbc2}
\frac{\partial g_{i,j}^\text{ct}\left[\mathbf{s}\right]}{\partial \mathbf{u}_j} = \pm \frac{1}{u_\text{ct}} \mathbf{l}_j = \pm \frac{1}{u_\text{ct}} \mathbf{f}_j,
\end{equation}
and the adjoint loads are therefore linearly dependent on \emph{non-corresponding} physical loads.
However, for $i,j = \{2,6\}$ and $i,j= \{4,8\}$ the adjoint load can \emph{not} be written as (a combination) of previously evaluated physical and/or adjoint loads and the corresponding systems of equations (\cref{eq:alamb}) need to be solved accordingly. Note, \emph{only} two solves are required as the adjoint loads for the constraints related to lower and upper bounds are linear dependent (these only show a sign difference).

Lastly, the adjoint loads corresponding to transmission constraint $g_{i,j}^\text{t}\left[\mathbf{s}\right]$ are given by
\begin{equation}
\label{eq:dbc3}
\frac{\partial g_{i,j}^\text{t}\left[\mathbf{s}\right]}{\partial \mathbf{u}_j} = \pm \left(\frac{J_k}{u_\text{t}} \mathbf{l}_i - \frac{1}{u_\text{t}} \mathbf{l}_j\right),
\end{equation}
which can all be written as a summation of the previous adjoint loads of $g_{i,j}^\text{in}\left[\mathbf{s}\right]$ (or physical loads $\mathbf{f}_6$ and $\mathbf{f}_8$ and $g_{i,j}^\text{ct}\left[\mathbf{s}\right]$. For such `combined' loads it can be particularly obscure to manually express them as a linear combination of previous physical and/or adjoint loads.

The problem analysis reveals that if no linear dependencies are taken into account 40 systems of equations need to be solved (of which 34 in the sensitivity analysis), as opposed the minimum of 8 when considering all linear dependencies (MLDs). 
That is, one may expect a maximum decrease of computational effort by 80\%. If only LDAP pairs are considered (this is generally the case), then 34 equations have to be solved. If, in addition to this, it is recognized that the adjoint loads of the constraints on lower and upper bounds only differ by a sign (and are thus linearly dependent), one still has to solve 20 systems of equations.

The results of the foregoing problem analysis are summarized in \cref{tab:my_label}, aiding in the detection of linear dependency between loads and calculation of states. Although manually finding all linear dependencies and their corresponding coefficients is achievable and yields significant savings, it is time-consuming, cumbersome, and error-prone. Moreover, it does not easily permit implementation in commercial software. In the following, we demonstrate how an LDAS, such as \cref{alg:axb}, provides the same result in an automated manner with negligible computational overhead.

\subsection{Verification by run-time experiment}
The following discusses a run-time measurement comparison between the LDAS and manual implementations considering LDAP pair and MLD detection. Our aim is to measure the gain of using an automatic LDAS as opposed to manual implementations, with a specific interest of the attained performance improvements across a range of number of DOFs $n$ for a single design iteration. 
All presented run-times are normalised with respect to an implementation without exploiting any linear dependencies. From the previous problem analysis, we found the number of solves required for each method: 40 for no detection, 34 considering LDAP, and 8 when including MLD, already hinting at potential performance improvements. 

In order to consider the influence of different types of solutions methods, we define the ratio $\chi$ as the ratio between the computational effort a solution method requires for preprocessing and the effort required for a solve. 
To capture a wide range of solution methods, we opt to compare two extremes:
\begin{itemize}
\item a high-$\chi$ solution method with predominant effort in the preprocessing; we opt here for a direct method, such as a Cholesky factorization \autocite{cholesky1924} with back-substitution, and
\item an a low-$\chi$ solution method with predominant effort in solving the equations. We opt here for an iterative solution process, such as an Incomplete Cholesky preconditioning with Conjugate Gradient \autocite{Saad2003}.
\end{itemize}

\begin{figure*}
\centering
\newlength\figH
\newlength\figW
\setlength{\figW}{0.3\textwidth}
\setlength{\figH}{0.15\textheight}
%
%
\begin{tikzpicture}

\begin{axis}[%
width=\figW,
height=\figH,
at={(6.5cm,0)},
scale only axis,
xmode=log,
xmin=1000,
xmax=100000,
xminorticks=true,
ymin=0,
ymax=1,
ylabel={$\hat{t}$},
ylabel near ticks,
ylabel style={rotate=-90},
xlabel={$n$},
axis background/.style={fill=white},
title style={font=\bfseries},
title={High-$\chi$ solution method},
legend style={at={(1,1)}, anchor=north west, legend cell align=left, align=left, fill=none, draw=none}
]
\addplot [color=blue, dashed, line width=2.0pt, mark size=2.0pt, mark=*, mark options={solid, blue}]
  table[row sep=crcr]{%
1000	0.883162423465075\\
2512	0.889196625065515\\
6310	0.891176280034821\\
15849	0.883447297330267\\
39811	0.902991876201297\\
100000	0.895701565557681\\
};
\addlegendentry{LDAP}

\addplot [color=black, dashed, line width=2.0pt, mark size=2.0pt, mark=*, mark options={solid, black}]
  table[row sep=crcr]{%
1000	0.386983337982082\\
2512	0.413973231308401\\
6310	0.420108414381568\\
15849	0.384368569040901\\
39811	0.438814808829267\\
100000	0.453668469461054\\
};
\addlegendentry{MLD}

\addplot [color=red, dashed, line width=2.0pt, mark size=2.0pt, mark=triangle, mark options={solid, red}]
  table[row sep=crcr]{%
1000	0.548339612116042\\
2512	0.547827685830616\\
6310	0.468406672763989\\
15849	0.411692934382429\\
39811	0.445812652561836\\
100000	0.469051013342174\\
};
\addlegendentry{LDAS}

\end{axis}

\begin{axis}[%
width=\figW,
height=\figH,
at={(0,0)},
scale only axis,
xmode=log,
xmin=1000,
xmax=100000,
xminorticks=true,
ymin=0,
ymax=1,
ylabel={$\hat{t}$},
ylabel near ticks,
ylabel style={rotate=-90},
xlabel={$n$},
axis background/.style={fill=white},
title style={font=\bfseries},
title={Low-$\chi$ solution method},
legend style={at={(1,1)}, anchor=north west, legend cell align=left, align=left, fill=none, draw=none}
]
\addplot [color=blue, dashed, line width=2.0pt, mark size=2.0pt, mark=*, mark options={solid, blue}]
  table[row sep=crcr]{%
1000	0.859270548607528\\
2512	0.855563486654185\\
6310	0.850153653991891\\
15849	0.850521964463755\\
39811	0.85052228161851\\
100000	0.843444979663469\\
};

\addplot [color=black, dashed, line width=2.0pt, mark size=2.0pt, mark=*, mark options={solid, black}]
  table[row sep=crcr]{%
1000	0.203305762501711\\
2512	0.200691113622673\\
6310	0.200721076091077\\
15849	0.200859190872439\\
39811	0.200912429661854\\
100000	0.199282646141453\\
};

\addplot [color=red, dashed, line width=2.0pt, mark size=2.0pt, mark=triangle, mark options={solid, red}]
  table[row sep=crcr]{%
1000	0.206831941211083\\
2512	0.213223963680587\\
6310	0.21240391500981\\
15849	0.203740170492545\\
39811	0.20532455586785\\
100000	0.203616133247433\\
};

\end{axis}

\end{tikzpicture}%
\caption{Normalized run-time $\hat{t}$ versus number of DOFs $n$ of three implementations: LDAP ($\bullet$), MLD by problem analysis ($\bullet$), and LDAS by automatic detection ($\blacktriangle$). The figures include both a high-$\chi$ and low-$\chi$ solution method to solve the system of equations related to the numerical example presented in \cref{sec:numerical}. For each of the six data points, the measurements are averaged over respectively 1000, 250, 64, 16, 4 and 1 repeated experiments on a high performance computing cluster to obtain a stable time measurement.}
\label{fig:run_time_results}
\end{figure*}
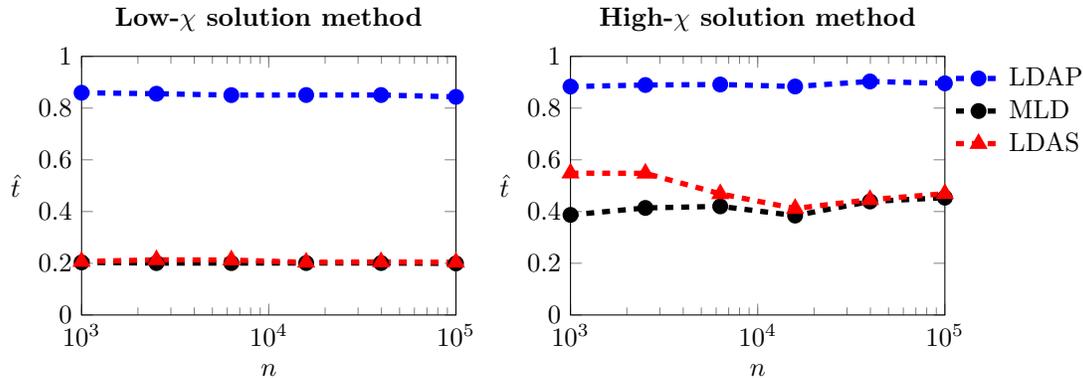

The presented experiments consider a moderate number of DOFs: small enough to highlight the change in performance as the number of DOFs is increased, while large enough to ensure the computational effort and run-time is dominated by preprocessing and solving. These aspects are therefore emphasized in the following analysis and other computational overhead is assumed negligible\footnote{Although very little computational overhead is present in the manual approaches, the required problem analysis (\cref{sec:probanalysis}) is time consuming and error-prone.}.
In all cases we reused the preprocessing information (factorization/preconditioner) when possible.
The results of this run-time experiment are shown in \cref{fig:run_time_results}. 
The figures show the normalized run-time $\hat{t}$, i.e. normalised with respect to the run-time required without any linear dependency detection, of the solves required for a single design iteration, both for high and low-$\chi$ methods.

For high-$\chi$ solution methods, the gains for LDAS and MLD converge toward each other, indicating the relative overhead of the LDAS decreases with problem size. It should be noted that the ideal normalised run time $\hat{t} = 0.2$ is not achieved for high-$\chi$ methods, since the chosen preprocessing is relatively expensive (or vice versa the solve is relatively cheap), thereby limiting the possible gains in run-time in this situation to $\hat{t} = 0.4$.
Clearly, the maximum achievable gain is higher for low-$\chi$ solution methods (the difference is fully defined by the difference in $\chi$), for which counting the number of linearly independent solves of the different schemes gives an accurate estimate of relative computational efficiency. 
That is, for the presented example, an 80\% reduction may indeed be expected using an LDAS with low-$\chi$ solution method.

Regardless of the solution method, taking into account only LDAP pairs is not computationally efficient compared to using LDAS, for this problem.
For both high-$\chi$ and low-$\chi$ solution methods the overhead of the LDAS is negligible for problems of moderate to large size.

\section*{Conclusions}
\label{sec:conclusions}
The computational effort required to solve a gradient-based structural optimisation problem in a nested analysis and design setting, is typically dominated by finding solutions to state equations. However, in real-world optimisation problems---that are typically \emph{compound}, i.e. they consider multiple combinations of physical loading conditions and a wide variety of response functions---many avoidable linear system solves are executed regardless. This paper presents proposes the use of linear dependency aware solvers, complementary to methods aiming to reduce the total number of design iterations, or the cost per solve, by effectively reducing the \emph{number} of solves per design iteration without compromising accuracy. The proposed concept leverages the linearity of the systems of equations---a trait present in many commonly considered topology optimisation problems---to automatically omit expensive solves if the solutions can be expressed as a linear combination of previously evaluated solutions for a given design iteration.

The present work present one such algorithm, that is simple, as illustrated by the provided supplementary Python and MATLAB implementations of \cref{alg:axb}, and can be integrated non-intrusively in existing optimisation software. Additionally, the concept does not restrict the use of other methods to reduce computational time per solve, such as parallel computing, approximation techniques, or model order reduction, which allows the user to focus on the design problem formulation and avoids laborious manual linearly dependency analysis altogether. Although the potential benefits of the proposed method hinges on the presence of linear dependencies of the problem at hand, it has been illustrated that the accompanying overhead is negligible, allowing the method to be applied freely, and achieving significant performance improvements when linear dependencies are abundant.

\section*{Replication of results}
\label{sec:ror}
A Python and MATLAB implementation of \cref{alg:axb,alg:gs} are available at GitHub: \url{https://github.com/artofscience/LDAS}.

\paragraph{Supplementary information}
This article is supplemented with numerical implementations, \textit{i.e.} a MATLAB and Python implementation of \cref{alg:gs} and \cref{alg:axb}, as well as media files related to the prototype model from \cref{fig:proto}. 

\paragraph{Declarations}
The authors declare that they have no conflict of interest.

\printbibliography

\appendix

\def\arraystretch{1.5}
\begin{table*}[]
\centering
\begin{tabular}{r|p{1cm}p{1cm}p{1cm}p{1cm}p{1cm}p{1cm}p{1cm}p{1cm}}
\toprule
Loads & 1 & 2 & 3 & 4 & 5 & 6 & 7 & 8\\
\midrule
$\mathbf{f}_1$ & $1$ &&&&&&&\\
$\mathbf{f}_3$ && & $1$ &&&&&\\
$\mathbf{f}_5$ &&&&& $1$ &&&\\
$\mathbf{f}_7$ & &&&&&& $1$ &\\
\midrule
$\mathbf{f}_6$ &&&&&& $1$ &&\\
$\mathbf{f}_8$ & &&&&&&& $1$\\
\midrule
$\frac{\partial f}{\partial \mathbf{u}_1}$ & $1$ &&&&&&&\\
$\frac{\partial f}{\partial \mathbf{u}_3}$ && & $1$ &&&&&\\
$\frac{\partial f}{\partial \mathbf{u}_5}$ &&&&& $1$ &&&\\
$\frac{\partial f}{\partial \mathbf{u}_7}$ & &&&&&& $1$ &\\
\midrule
$\frac{\partial g^\text{in}_{6,6}}{\partial \mathbf{u}_6}$ & &&&&& $\frac{1}{u_\text{in}}$&&\\
$\frac{\partial g^\text{in}_{8,8}}{\partial \mathbf{u}_8}$ & &&&&&&& $\frac{1}{u_\text{in}}$\\
\midrule
$\frac{\partial g^\text{ct}_{1,6}}{\partial \mathbf{u}_6}$ & $\frac{1}{u_\text{ct}}$ &&&&&&&\\
$\frac{\partial g^\text{ct}_{1,6}}{\partial \mathbf{u}_6}$ & $-\frac{1}{u_\text{ct}}$ &&&&&&&\\
$\frac{\partial g^\text{ct}_{2,6}}{\partial \mathbf{u}_6}$ && $\frac{1}{u_\text{ct}}$ &&&&&&\\
$\frac{\partial g^\text{ct}_{2,6}}{\partial \mathbf{u}_6}$ && $-\frac{1}{u_\text{ct}}$ &&&&&&\\
$\vdots$\\
$\frac{\partial g^\text{ct}_{8,6}}{\partial \mathbf{u}_6}$ &&&&&&&& $\frac{1}{u_\text{ct}}$ \\
$\frac{\partial g^\text{ct}_{8,6}}{\partial \mathbf{u}_6}$ &&&&&&&& $-\frac{1}{u_\text{ct}}$ \\
\midrule
$\frac{\partial g^\text{ct}_{1,8}}{\partial \mathbf{u}_8}$ & $\frac{1}{u_\text{ct}}$ &&&&&&&\\
$\frac{\partial g^\text{ct}_{1,8}}{\partial \mathbf{u}_8}$ & $-\frac{1}{u_\text{ct}}$ &&&&&&&\\
$\frac{\partial g^\text{ct}_{3,8}}{\partial \mathbf{u}_8}$ &&& $\frac{1}{u_\text{ct}}$ &&&&&\\
$\frac{\partial g^\text{ct}_{3,8}}{\partial \mathbf{u}_8}$ &&& $-\frac{1}{u_\text{ct}}$ &&&&&\\
$\vdots$\\
$\frac{\partial g^\text{ct}_{7,8}}{\partial \mathbf{u}_8}$ &&&&&&& $\frac{1}{u_\text{ct}}$ &\\
$\frac{\partial g^\text{ct}_{7,8}}{\partial \mathbf{u}_8}$ &&&&&&& $-\frac{1}{u_\text{ct}}$ &\\
\midrule
$\frac{\partial g^\text{t}_{4,6}}{\partial \mathbf{u}_6}$ &&&& $\frac{J_{4,6}}{u_\text{t}}$ &&$-\frac{1}{u_\text{t}}$&&\\
$\frac{\partial g^\text{t}_{4,6}}{\partial \mathbf{u}_6}$ &&&& $-\frac{J_{4,6}}{u_\text{t}}$ &&$\frac{1}{u_\text{t}}$&&\\
\midrule
$\frac{\partial g^\text{t}_{2,8}}{\partial \mathbf{u}_8}$ && $\frac{J_{2,8}}{u_\text{t}}$ &&&&&&$-\frac{1}{u_\text{t}}$\\
$\frac{\partial g^\text{t}_{2,8}}{\partial \mathbf{u}_8}$ && $-\frac{J_{2,8}}{u_\text{t}}$ &&&&&&$\frac{1}{u_\text{t}}$\\
\bottomrule
\end{tabular}
\caption{Result of the problem analysis (\cref{sec:probanalysis}); relation between loads and DOF of interest. The horizontal axis states the eight DOFs of interest, and the vertical axis the physical and adjoint loads, respectively.}
\label{tab:my_label}
\end{table*}

\end{document}